\documentclass[aps,prd,twocolumn,10pt]{revtex4}
\usepackage[colorlinks,linkcolor={red},citecolor={blue}]{hyperref}
\usepackage{eurosym}
\usepackage{graphicx}
\usepackage{amsmath}
\usepackage{amssymb}
\usepackage{amsfonts}
\usepackage{bm}
\usepackage{float}
\usepackage{color}
\usepackage{multirow}
\usepackage{hhline}

\def\be{\begin{equation}}
\def\ee{\end{equation}}
\def\bea{\begin{eqnarray}}
\def\eea{\end{eqnarray}}

\newcommand{\mc}[1]{\mathcal{#1}}

\newcommand{\f}[2]{\frac{#1}{#2}}

\begin{document}

\title{Matter perturbations in Einstein dark energy model}
\author{Zahra Haghani}
\email{z.haghani@du.ac.ir}
\affiliation{School of Physics, Damghan University, Damghan 41167-36716, Iran.}

\date{\today }

\begin{abstract}
We consider the growth rate of matter perturbations in the Einstein dark energy theory. The theory consists of the Einstien-Hilbert Lagrangian plus the trace of the energy momentum tensor, coupled non-minimally to a dynamical vector field. We will show that the theory has three fixed points corresponding to the dust, radiation and de Sitter universes. Due to the present of trace of the energy-momentum tensor in the Lagrangian, the fixed points occurs in different locations compared to $\Lambda$CDM theory. We will analyze the theory with and without  cosmological constant. We will fit the model parameters using two independent data sets corresponding to the Hubble parameter $H$ and also $f\sigma_8$. The theory is then shown to be consistent with observational data.
\end{abstract}

\maketitle

\section{Introduction}
After introducing the general theory of relativity \cite{Ein1}, Einstein in 1919 tried to formulate the elementary particles interactions through general relativity \cite{Ein2}. He then considered an electromagnetic like matter field with energy-momentum tensor $S_{\mu\nu}$ representing elementary particles \cite{Ein2}. Since this tensor is trace-less, he modified the Einstein equation to describe the new interaction as
\begin{align}\label{pareq}
R_{\mu\nu}-\f14Rg_{\mu\nu}=\kappa^2 S_{\mu\nu}.
\end{align}
Moreover, Einstein assumed that elementary particle field satisfies Maxwell's equation.
So taking the covariant derivative of the above equation yields
\begin{align*}
\f14 \nabla_\mu R=-\f{\kappa^2}{c} F_{\mu\nu}j^\nu,
\end{align*}
where $j^\nu$ is the electric current. It is evident that in the volume outside the elementary particles and in the absence of charges	  we obtain from the above equation that $$R=R_0=constant.$$
Einstein, also assumed that the standard equation of general relativity still holds, so we also have
\begin{align}\label{GReq}
R_{\mu\nu}-\f12Rg_{\mu\nu}+\Lambda g_{\mu\nu}=\kappa^2T_{\mu\nu},
\end{align}
where $T_{\mu\nu}$ is the energy-momentum tensor of the baryonic matter. In vacuum, from the above equation we obtain
$$R=4\Lambda,$$ which by comparison with the result from the elementary particles equation we obtain
\begin{align}\label{lameq}
\Lambda=\f{R_0}{4}.
\end{align}
This shows that the cosmological constant can be considered as an integration constant from the elementary particles field equation \cite{Ein2}. By using equations \eqref{pareq}, \eqref{GReq} and \eqref{lameq} one can obtain the so-called matter-geometry symmetric Einstein equation
\begin{align}
R_{\mu\nu}-\f14Rg_{\mu\nu}=\kappa^2\left(T_{\mu\nu}-\f14Tg_{\mu\nu}\right),\label{eq01}
\end{align}
and
$$S_{\mu\nu}=T_{\mu\nu}-\f14Tg_{\mu\nu},$$
where $S_{\mu\nu}$ as was mentioned before is the energy-momentum tensor of the elementary particles in the form of electromagnetic fields. For a dust universe we obtain a very interesting result from the above equations. In this case $T^0_0=-\rho$ and then we obtain $S^0_0=-3/4\rho$. This means that the Einstein model predicts that the ingredient of matter in the universe is $75\%$ electromagnetic and $25\%$ gravitational. This theory has also been introduced as the unimodular gravity, which can be formulated in different ways. In this theory the determinant of  metric tensor is constrained to be a number or scalar density. As a consequence the cosmological constant appears as a  constant of integration. The action of unimodular gravity with a fixed metric determinant i.e. $\sqrt{-g}=\epsilon_0$ \cite{ug1} is 
\begin{align}
S_{UG}=\int d^4x \left[\sqrt{-g}\kappa^2 R-\lambda\left(\sqrt{-g}-\epsilon_0\right)\right]+S_m,
\end{align}  
where $\lambda$ is the Lagrange multiplier and $S_m$ is the action of matter fields. It should be noted that the action $S_{UG}$ is invariant under a restricted group of diffeomorphisms  in which the determinant of  metric tensor  is unchanged. The field equation of this model is the same as Einstein-Hilbert equations with cosmological constant, together with the constraint  $\sqrt{-g}=\epsilon_0$. By assuming the conservation of the energy momentum tensor one can obtain  $\nabla_\mu \lambda=0$. Henneaux and Teitelboin introduced a unimodular action which is fully diffeomorphism invariant \cite{HTUG}. The action of this model is as follow
 \begin{align}
 S_{HT}=\int d^4x \left[\sqrt{-g}\kappa^2 R-\lambda\left(\sqrt{-g}-\partial_\mu\tau^\mu\right)\right]+S_m,
 \end{align}  
where $\tau^\mu$ is a vector density. The equation of motion for the metric tensor is the same as Einstein-Hilbert field equations with cosmological constant but the determinant of the metric is constrained to be $\sqrt{-g}=\partial_\mu \tau^\mu$. The actions $S_{UG}$ and $S_{HT}$ are classically equivalent and by  a change of coordinates they are related to each other. There is an alternative action which is fully diffeomorphism invariant \cite{dug} as

\begin{align}
S_{DUG}=\int d^4x \sqrt{-g}\left[\kappa^2 R-\lambda+V^\mu\nabla_\mu \lambda\right]+S_m,
\end{align}
where the vector field $V^\mu$ is the Lagrange multiplier to keep constraint $\nabla_\mu\lambda=0$. In this action the constraint on the value of the metric determinant is replaced by $\nabla_\mu V^\mu=1$. By integrating by part of the action $S_{HT}$, one can easily  show that the two actions $S_{HT}$ and $S_{DUG}$ are equivalent at classical level. In \cite{ug1} the authors use the path integral to show the differences between the above mentioned models of unimodular gravity at quantum level. However, the introduction of unimodular gravity in a more general way has been reviewed in \cite{revug}.

Rastall \cite{rastall} is the first person who considered the modified equation \eqref{eq01} with the baryonic energy-momentum tensor instead of elementary $S_{\mu\nu}$ tensor. He also assumed a general coupling for the Ricci scalar
$$R_{\mu\nu}+\lambda Rg_{\mu\nu}=\kappa^2T_{\mu\nu}.$$
This equation predicts that the energy-momentum tensor is no longer conserved and there is a chance to transform directly to geometry $\nabla_\nu T^{\mu\nu}\propto\nabla_\nu R$.

Rastall theory have been investigated vastly in the literature \cite{ras1}. Also, many generalizations and modifications of the idea has been proposed. One of the most interesting of them is to consider a theory containing a non-minimal coupling between matter Lagrangian and geometry \cite{frlm1}. The action can then be written in the form
\begin{align}
S=\int \sqrt{-g}d^4x\left(\kappa^2R+f(R,\mc{L}_m)+\mc{L}_m\right),
\end{align}
where $\mc{L}_m$ is the matter Lagrangian. This theory has a general property of the Rastall theory which is the non-conservation of the energy-momentum tensor. This  causes the matter fields to be converted directly to geometry. Cosmological implications of this theory is vastly investigated \cite{frlm1}. Other generalizations of this idea includes non-minimal coupling between matter energy-momentum tensor and geometry such as $f(R,T)$ \cite{frt1}, $f(R,T,R_{\mu\nu}T^{\mu\nu})$ \cite{frtmu1} gravity theories. Also, one can consider non-standard interactions between matter fields such as $f(T,T_{\mu\nu}T^{\mu\nu})$ theories where $T$ is the trace of energy-momentum tensor \cite{emsg}, or derivative matter couplings \cite{derivative} where one considers interactions of the form $\nabla_\mu\mc{L}_m\nabla^\mu\mc{L}_m$.

In this paper, we are going to consider the cosmological implications of the Einstein dark energy theory introduced in \cite{EDE}, where the spirit of Einstein idea is putted together with the properties of Rastall gravity.  In this way, we will consider an Einstein-Hilbert action coupled with a dark energy vector field denoted as $\Lambda_\mu$ which is minimally coupled to geometry, but non-minimally coupled to the baryonic energy-momentum tensor through the interaction of the form $\mc{L}_m\Lambda_\mu\Lambda^\mu$.  We also considered the Rastall theory by adding a term proportional to the trace of the energy-momentum tensor. It should be noted that the present theory is different from the unimodular  gravity since in the Einstein dark energy theory the vector field does not constraint the metric field and it is an independent dynamical field. Also, since the vector field has a non-minimal coupling with matter Lagrangian, the matter fields do not conserve in this theory as opposed to the unimodular gravity.

In \cite{EDE}, the authors have shown that the theory can describe the late time accelerated expansion of the universe. However, the theory was not fully satisfactory with recent observational data. In this paper, by adding a non-minimal imteraction term between matter fields and the dark energy vector field, we will analyze the cosmological implications in more details and show that the modified theory is in fact capable of explaining the recent observational data in both background and first order perturbation levels. We also estimate the values of the model parameters to obtain the best fit of the theory with experiments by two sets of data corresponding to the Hubble \cite{hubble} and $f\sigma_8$ \cite{fsigma8} functions. We have also considered the dynamical system analysis of the model and show that the theory has three fixed points. The theory we are considering does not have a conservation of the energy-momentum tensor. This implies that the behavior of the energy density in dust/radiation dominated universes are not the same as in general relativity. As a result two of the fixed points of the Einstein dark energy model are the would be dust and radiation dominated fixed points which now behaves differently due the the presence of non standard matter couplings in the theory. The third fixed point corresponds to the de Sitter expansion of the universe. Similar to general relativity, this fixed point is stable. So, the theory can in principle explain the thermal history of the universe. Also, we will analyze the theory at the perturbative level and obtain the evolution equation corresponding to the growth rate of matter perturbations.

The structure of the paper is as follows. In the next section we review the Einstein dark energy model and obtain main dynamical equations of the model. In section \ref{cos}, we consider the background cosmological implications of the model and in section \ref{dyna} we investigate its dynamical analysis. In section \ref{pert}, the matter scalar perturbations on top of flat FRW universe is considered and at the end we will conclude the paper.

\section{The model}\label{sect2}

In this section we present the field equations for the Einstein dark energy model, and derive some of its basic theoretical consequences \cite{EDE}. Let us assume that the universe is filled with a cosmological dark energy vector field $\Lambda _{\mu }\left( x^{\nu
}\right) $. We define the dark energy strength tensor as
\begin{equation}
C_{\mu \nu }=\nabla _{\mu }\Lambda _{\nu }-\nabla _{\nu }\Lambda _{\mu }.
\end{equation}%
The dark energy strength tensor identically satisfies the Maxwell type
equations
\begin{equation}
\nabla _{\lambda }C_{\mu \nu }+\nabla _{\mu }C_{\nu \lambda }+\nabla _{\nu
}C_{\lambda \mu }=0.  \label{13}
\end{equation}%
We define the energy-momentum tensor $T_{\mu \nu}$ of the baryonic matter fields as
\be
T_{\mu \nu}=-\frac{2}{\sqrt{-g}} \frac{\partial \left( \sqrt{-g} \mathcal{L}_{m} \right)}{%
\partial g^{\mu\nu}},
\ee
where  $\mathcal{L}_{m}$ is the Lagrangian of the total (ordinary baryonic plus dark) matter.
In the following, by $T$ we denote the trace of the matter energy-momentum tensor.

The Einstein dark
energy model is described by the following action
\begin{align}
S =\int d\,^4x\sqrt{-g}\,\Bigg[&\kappa^2\left( 1-\beta_1 \right) R+%
\frac{\beta_2 }{2}\,T-\frac{1}{4}C_{\mu \nu }C^{\mu \nu }  \notag  \label{s1}
\\
&+\alpha\,\mathcal{L}_m\,\Lambda _{\mu }\Lambda^\mu +V(\Lambda^2)+\mathcal{L}_{m} \Bigg] ,
\end{align}%
where $\beta_1$, $\beta_2$ are two arbitrary dimensionless constants and $\alpha $ is a coupling constant with mass dimension $M^{-2}$ representing the interaction between matter and the dark energy vector $\Lambda_\mu$. Also, the potential term $V$ is an arbitrary function of $\Lambda^2=\Lambda_\mu\Lambda^\mu$. In this paper, we will consider the constant potential corresponding to the cosmological constant, and also a power-law case.

The energy-momentum tensor $S_{\mu \nu }$ of the dark energy field can be obtained by varying its kinetic term with respect to the metric, which gives
\begin{equation}  \label{23}
S_{\mu \nu }=C_{\mu \alpha }C_{\nu }^{~\alpha }-\frac{1}{4}g_{\mu \nu
}C_{\alpha \beta }C^{\alpha \beta },
\end{equation}
with the property $S_{\mu}^{\mu}=0$.

By varying the gravitational
action with respect to the metric tensor, it follows that the cosmological evolution of the universe
in the presence of a vector type dark energy is described by the generalized
Einstein gravitational field equations,
\begin{align}\label{eq1}
&\kappa^2(1-\beta_1)G_{\mu\nu}-\f12S_{\mu \nu}
\nonumber\\
&+\alpha\left(\mathcal{L}_m\Lambda_\mu\Lambda_\nu-\frac{1}{2}\Lambda^2T_{\mu\nu}\right)-\frac{1}{2}g_{\mu\nu}V+\Lambda_\mu\Lambda_\nu V^\prime \nonumber\\
&=\f12(1+\beta_2)T_{\mu\nu}-\f12\beta_2\,\left(\mathcal{L}_m-\frac{1}{2}T\right)g_{\mu\nu},
\end{align}
where prime denotes derivative with respect to the argument. 

By varying the action \eqref{s1} with respect to the vector potential, we obtain the equation
\begin{equation}
\nabla _{\nu }C^{\mu\nu }=2\Lambda^\mu(V^\prime+\alpha \mathcal{L}_m).\label{14}
\end{equation}
It should be noted that due to the non-minimal coupling between matter and geometry, the matter field is the source for the dark energy vector field.

By taking the divergence of the metric field equation \eqref{eq1} and using equation \eqref{14} one obtains the  conservation equation of the energy-momentum tensor as
\begin{align}\label{cons1}
\nabla^\mu T_{\mu\nu}&=\f{1}{1+\beta_2+\alpha\Lambda^2}\Big[\beta_2\nabla_\nu\left(\mathcal{L}_m-\f12T\right)\nonumber\\&\quad\qquad+\alpha(\mathcal{L}_mg_{\mu\nu}-T_{\mu\nu})\nabla^\mu(\Lambda^2)\Big].
\end{align}
It can be seen from the above equation that there are two sources for the non-conservation of the energy-momentum tensor. The first one is due to the presence of the trace of the energy-momentum tensor $T$ and the second one comes from the non-minimal coupling between matter and the dark energy vector potential in the action. In the case $\alpha=0=\beta_2$ the energy-momentum tensor becomes conserved. We will defined a vector field
\begin{align}\label{rest}
f_\nu\equiv\f{1}{1+\beta_2+\alpha\Lambda^2}&\Big[\beta_2\nabla_\nu\left(\mathcal{L}_m-\f12T\right)\nonumber\\&+\alpha(\mathcal{L}_mg_{\mu\nu}-T_{\mu\nu})\nabla^\mu(\Lambda^2)\Big],
\end{align}
which is the right-hand side of equation \eqref{cons1} and represents the amount of non-conservation of the energy-momentum tensor. In the case of $f_\nu=0$, the energy-momentum tensor becomes conserved.
\section{Cosmological implications}\label{cos}
\begin{figure*}
	\includegraphics[scale=0.55]{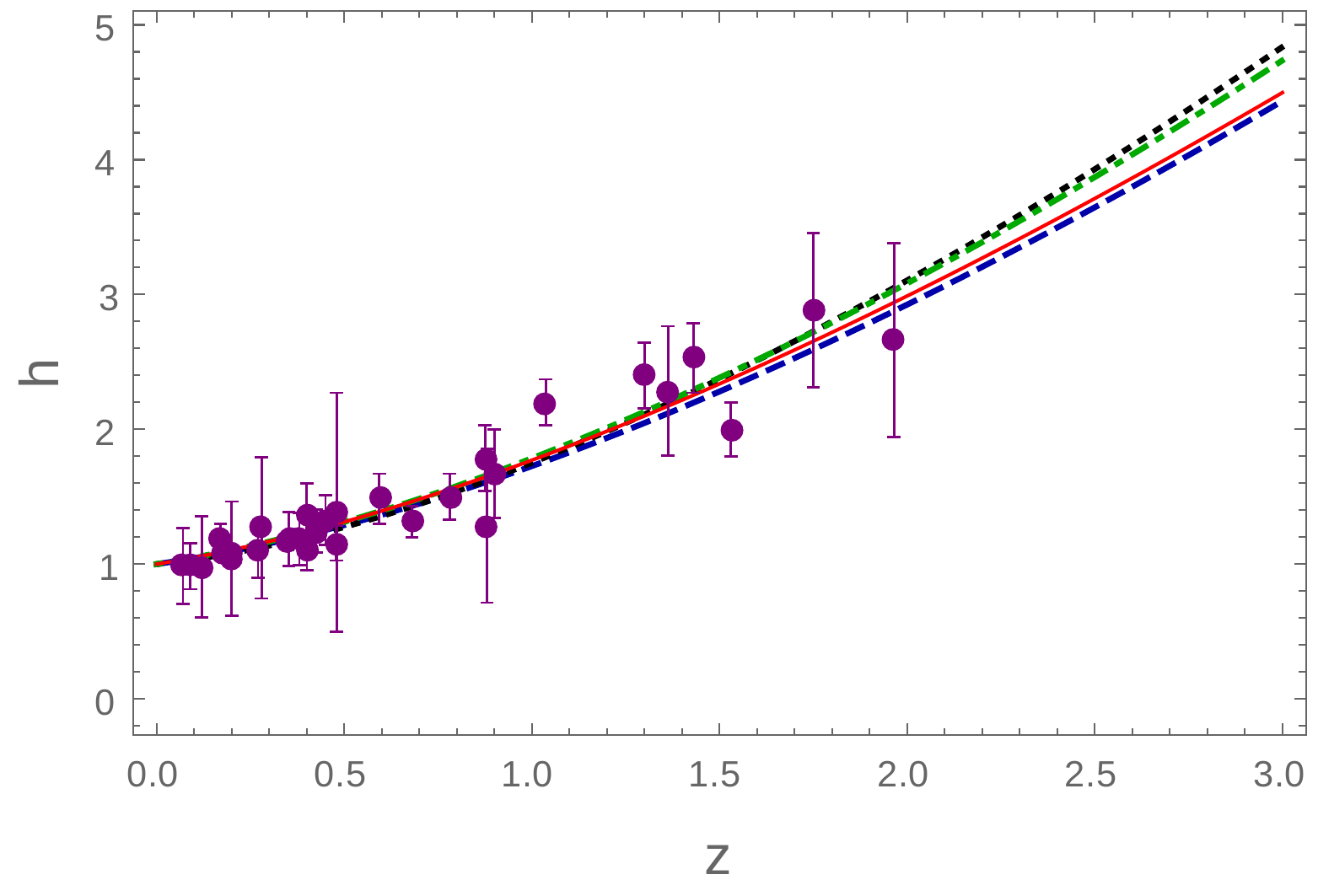}
	\includegraphics[scale=0.57]{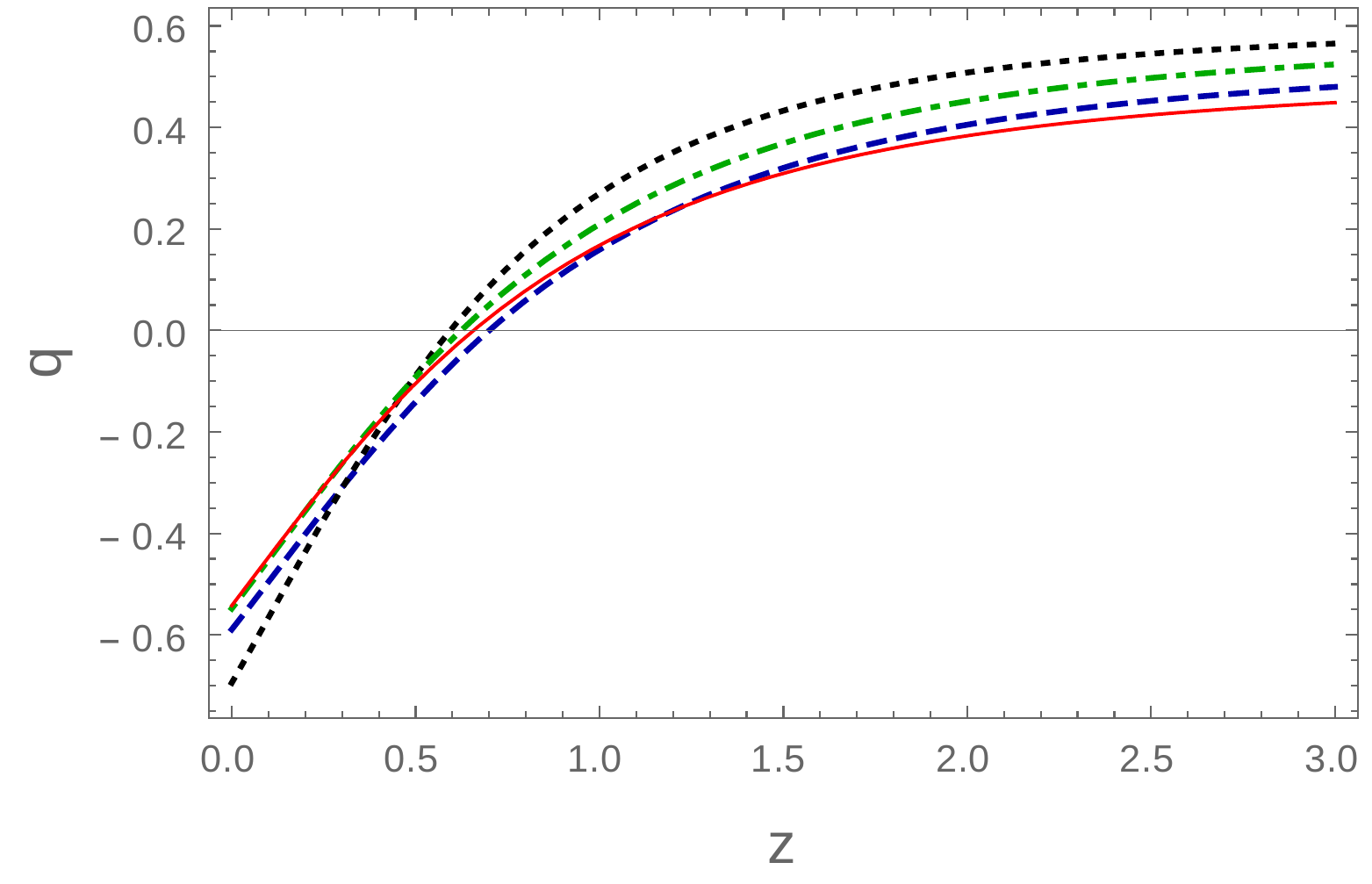}
	\caption{The Hubble and deceleration parameters as a function of the redshift $z$ for three different values for the constant $\eta=-0.02$ (dot-dahsed), $\eta=0$ (dashed) and $\eta=0.1$ (dotted). We use the best fit values for the model parameters presented in table \eqref{tab2}. For the Hubble parameter, we have also plotted te experimental data together with their errors \cite{hubble}. The $\Lambda$CDM curve is depicted as a red solid curve.\label{fig1}}
\end{figure*}
Let us consider the flat FRW universe with conformal time $t$
\begin{align}
ds^2=a^2\eta_{\mu\nu}dx^\mu dx^\nu,
\end{align}
where $a=a(t)$ is the scale factor. The Hubble parameter can be defined as $H=\dot{a}/a$, where dot represents derivative with respect to the conformal time.

For the dark energy vector field, we assume that only the temporal component is non-vanishing
\begin{align}
\Lambda_\mu= a\big[\Lambda_0(t),\vec{0}\big],
\end{align}
which is dictated by the isotropy and homogeneity of the FRW space time in the Cartesian coordinates. 
We also assume that the universe is filled with a perfect fluid with Lagrangian density $\mathcal{L}_m=-\rho$ and energy-momentum tensor  
\begin{equation}
T_{\mu \nu }=\left( \rho+p\right) u_{\mu}u_{\nu }+pg_{\mu \nu },  \label{18}
\end{equation}%
where $\rho$ is the energy density and $p$ is the thermodynamics pressure.

The Friedmann  and Raychaudhuri equations can be obtained from \eqref{eq1} as
\begin{align}\label{frid1}
\f{6\kappa^2}{a^2}(1-\beta_1)H^2&=(1+\alpha\Lambda_0^2)\rho\nonumber\\&+\f12\beta_2(\rho-3p)-(V+2\Lambda_0^2V^\prime),
\end{align}
\begin{align}\label{frid2}
\f{2\kappa^2\left(1-\beta _1\right)}{a^2}&\left( H^2+2 \dot{H}\right)\nonumber\\&=(\alpha\Lambda_0^2-1)p-\f12\beta_2(\rho+5p)-V.\end{align}
The field equation of the vector field is
\begin{align}\label{l0}
V^\prime=\alpha\rho.
\end{align}
From the above equation, one can see that in the case of a constant potential $V=const.$, the non-minimal coupling between the matter and the dark energy vector field should vanish.

The non-conservation equation of the matter field \eqref{cons1} is reduced to
\begin{align}\label{nc0}
\bigg(1&+\f12\beta_2-\alpha\Lambda_0^2\bigg)\dot\rho-\f32\beta_2\dot{p}\nonumber\\&+3(1+\beta_2-\alpha\Lambda_0^2)H(\rho+p)=0.
\end{align}
Now, let us assume a specific form of the potential as
\begin{align}
V(\Lambda^2)=-\beta_3\left(-\f{\Lambda^2}{\kappa^2}\right)^\eta,
\end{align}
where $\eta$ is a dimensionless constant and $\beta_3$ is a constant with mass dimension $M^4$.

In the case $\eta=0$, one has $V=-\beta_3$ which mimics the cosmological constant. So, we define the modified cosmological constant in this model as $\lambda=\beta_3/2\kappa^2$. In the case of $\eta=0$, $\lambda$ coincides with the standard cosmological constant. However, the value of $\lambda$ will differ from the cosmological constant for $\eta\neq0$.

Let us assume that the universe is filled with pressure-less dust and radiation. The energy density and pressure becomes
\begin{align}
\rho=\rho_m+\rho_r,\qquad p=\f13\rho_r,
\end{align}
where $m/r$ denotes $dust/radiation$ respectively.

Defining the following dimensionless parameters
\begin{align}
&\tau=H_0 t,\quad H=H_0 h, \quad \bar{\rho}_i=\f{\rho_i}{6\kappa^2H_0^2},\nonumber\\&
\beta_3=6\kappa^2H_0^2\Omega_\lambda,\quad \alpha_1=\alpha\kappa^2,\quad \Lambda_0=\kappa\Lambda_1,
\end{align}
where $H_0$ is the current Hubble parameter, one can write the metric field equations as
\begin{align}\label{frid3}
(1-\beta_1)h^2&=a^2\Bigg[\Omega_\lambda+\left(1+\f12\beta_2\right)\bar\rho_m\nonumber\\&+\alpha_1\Lambda_1^2(\bar\rho_m+\bar\rho_r)+(2\eta-1)\Omega_\lambda\Lambda_1^{2\eta}\Bigg],
\end{align}
\begin{align}
(1-\beta_1)(h^2+2h^\prime)&=-a^2\Bigg[\f32\beta_2\bar\rho_m-3\Omega_\lambda\Lambda_1^{2\eta}\nonumber\\&+\big(1+4\beta_2-\alpha_1\Lambda_1^2\big)\bar\rho_r\Bigg],
\end{align}
where prime here denotes derivative with respect to the dimensionless time $\tau$.
The vector field equation can be written as
\begin{align}\label{vec1}
\eta\Omega_\lambda\Lambda_1^{2(\eta-1)}=\alpha_1(\bar\rho_m+\bar\rho_m).
\end{align}
As we have discussed before, in the case of $\eta=0$ which corresponds to the case of cosmological constant, one should impose $\alpha_1=0$. So $\Lambda_1$ does not contribute to the background cosmological evolution of the universe. In the case $\eta\neq0$, one can obtain the value of $\Lambda_1$ from the vector field equation \eqref{vec1} as
\begin{align}\label{vec2}
\Lambda_1=\left[\f{\alpha_1(\bar\rho_m+\bar\rho_r)}{\eta\Omega_\lambda}\right]^{\f{1}{2(\eta-1)}}.
\end{align}
The baryonic matter conservation equation can be written in dimensionless coordinates as
\begin{align}\label{cons2}
\Big[(2&+\beta_2-2\alpha_1\Lambda_1^2)\bar\rho_m^\prime+6(1+\beta_2-\alpha_1\Lambda_1^2)h\bar\rho_m\Big]\nonumber\\&+\Big[2(1-\alpha_1\Lambda_1^2)\bar\rho_r^\prime+8(1+\beta_2-\alpha_1\Lambda_1^2)h\bar\rho_r\Big]=0.
\end{align}
From the structure of the above equation, we will assume that each bracket in \eqref{cons2} vanishes independently and as a result we have two  conservation equations for dust and radiation as
\begin{align}\label{con1}
\bar\rho_m^\prime+6h\left[\f{1+\beta_2-\alpha_1\Lambda_1^2}{2+\beta_2-2\alpha_1\Lambda_1^2}\right]\bar\rho_m=0,
\end{align}
\begin{align}\label{con2}
\bar\rho_r^\prime+4h\left[\f{1+\beta_2-\alpha_1\Lambda_1^2}{1-\alpha_1\Lambda_1^2}\right]\bar\rho_r=0.
\end{align}
It should be noted that in the case $\beta_2=0=\alpha_1$, the above equations becomes standard conservation equations for dust and radiation.

Let us define redshift parameter as
\begin{align}
1+z=\f1a.
\end{align}
Derivatives with respect to $\tau$ can be converted to the derivatives with respect to the redshift as
\begin{align}
\f{d}{d\tau}=-(1+z)h(z)\f{d}{dz}.
\end{align}
Using equations \eqref{frid3} and \eqref{vec1} and transforming to the redshift coordinates, one can obtain the dimensionless Hubble parameter as
\begin{align}\label{frid5}
h(z)&=\f{1}{\sqrt{1-\beta}}(1+z)^{-1}\Bigg[\left(1+\f12\beta_2\right)\bar\rho_m(z)+\bar\rho_r(z)\nonumber\\&+(1-\eta)\left[\f{1}{\Omega_\lambda}\left(\f{\alpha_1(\bar\rho_m(z)+\bar\rho_r(z))}{\eta}\right)^\eta\right]^{\f{1}{\eta-1}}\Bigg]^{1/2}.
\end{align}
The deceleration parameter in terms of redshift can be written as
\begin{align}
q=\left(1+z\right)\f{d \ln h}{dz}.
\end{align}
Noting that $h(0)=1$, $\bar\rho_m(0)=\Omega_{m0}=0.305$ and $\bar\rho_r(0)=\Omega_{r0}=5.3\times 10^{-5}$ in redshift coordinates \cite{plank}, one can see from equation \eqref{frid5} that the modified cosmological constant density parameter can be expressed by other parameters as
\begin{align}
\Omega_\lambda=\left(\f{\eta}{\alpha_1}\right)^\eta\left[\f{2(\eta-1)(\Omega_{m0}+\Omega_{r0})^{\f{\eta}{\eta-1}}}{2(\beta_1-1)+(2+\beta_2\Omega_{m0}+2\Omega_{r0})}\right]^{\eta-1}.
\end{align}
Also, we should note that on top of FRW space-time, the temporal component of the vector field $f_\nu$ is non-vanishing and in dimensionless coordinates is given by
\begin{align}\label{rest1}
f_0=\f{3\beta_2(1+z)h(\bar\rho_m^\prime+2\bar\rho_r^\prime)
}{1+\beta_2-\alpha_1\Lambda_1^2}.
\end{align}
In figure \eqref{fig1} we have plotted the evolution of the Hubble parameter $h$ together with the deceleration parameter $q$ as a function of redshift.
We have assumed three different values for the constant $\eta=-0.02,0,0.1$. It should be emphasized that in section \ref{pert}, we will obtain the best fit values of the model parameters $\beta_1$ and $\beta_2$ using two independent data sets of Hubble parameter and $f\sigma_8$. The best fit values of the parameters, together with their $1\sigma$ and $2\sigma$ confidence intervals are shown in table \eqref{tab1}, for three values of $\eta=-0.02,0,0.1$. In the figure \eqref{fig1}, we have used the best fit values. It can be shown from the figure that the Einstien dark energy model can explain the observational data on the Hubble parameter very well. In the case $\eta\neq0$, one can see that the Hubble parameter becomes larger than $\Lambda$CDM value for redshifts greater than $z\sim1.5$. In the case of cosmological constant $V(\Lambda^2)=-\beta_3$ however, the Hubble parameter is very close to the $\Lambda$CDM curve. However, the EDE prediction of the Hubble parameter for redshifts $z>1.5$ is a little smaller than $\Lambda$CDM value. This shows that the size of the universe is smaller for non-vanishing values of $\eta$. The evolution of the deceleration parameter shows that the universe have more deceleration compared to the $\Lambda$CDM model at redshift larger than $z\sim1.5$. It should be mentioned that the Einstein dark energy model with a cosmological constant behaves a little different from the other cases with non-vanishing $\eta$. This is due to the fact that in the case of $\eta=0$, the non-minimal coupling between matter and dark energy vector field vanishes, which makes the universe to gain more acceleration.

\begin{figure}
	\includegraphics[scale=0.5]{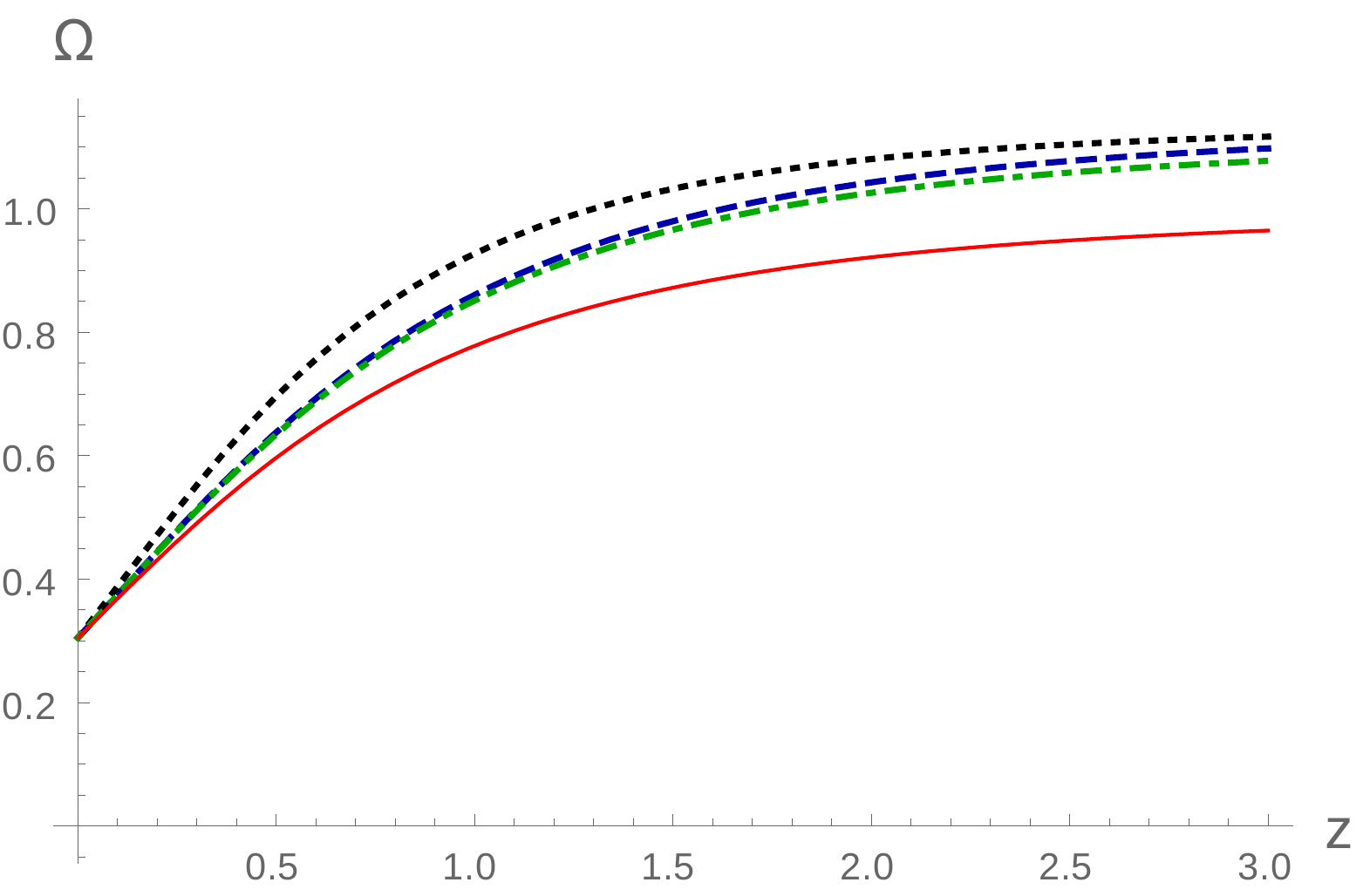}
	\caption{The evolution of the matter density abundance as a function of the redshift $z$ for three different values for the constant $\eta=-0.02$ (dot-dahsed), $\eta=0$ (dashed) and $\eta=0.1$ (dotted). We use the best fit values for the model parameters presented in \eqref{tab2}. The $\Lambda$CDM curve is depicted as a red curve.\label{figomega}}
\end{figure}

\begin{figure}
	\includegraphics[scale=0.5]{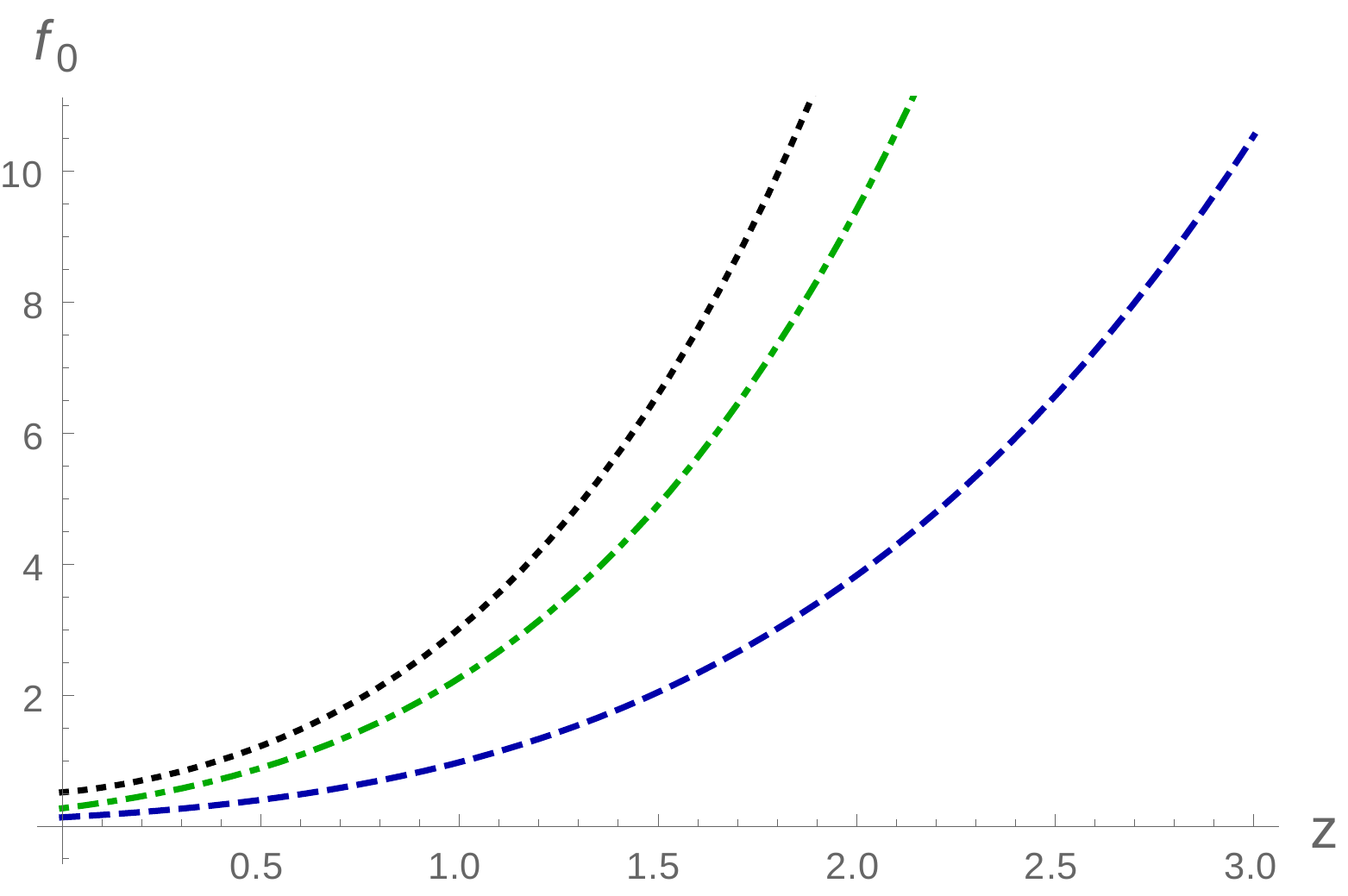}
	\caption{The evolution of the non-conservation of the energy-momentum tensor $f_0$ as a function of the redshift $z$ for three different values for the constant $\eta=-0.02$ (dot-dahsed), $\eta=0$ (dashed) and $\eta=0.1$ (dotted). We use the best fit values for the model parameters presented in table \eqref{tab2}.\label{figrest}}
\end{figure}

In figure \eqref{figomega}, we have plotted the evolution of the density abundance $\Omega_m$, defined as
$$\Omega_m=\f{\bar\rho_m a^2}{h^2}.$$
As can be seen from the figure, the baryonic matter density becomes larger than the conservative $\Lambda$CDM model. The difference can be seen as an amount of non-conservation of the energy momentum tensor. This can also be seen from equation \eqref{rest1} which we have plotted in figure \eqref{figrest}.
For redshifts greater than $z>0.2$ the non-conservation of the energy-momentum tensor $f_0$ becomes non-zero which causes the matter density abundance to behave differently from the $\Lambda$CDM theory.

\begin{figure}
	\includegraphics[scale=0.5]{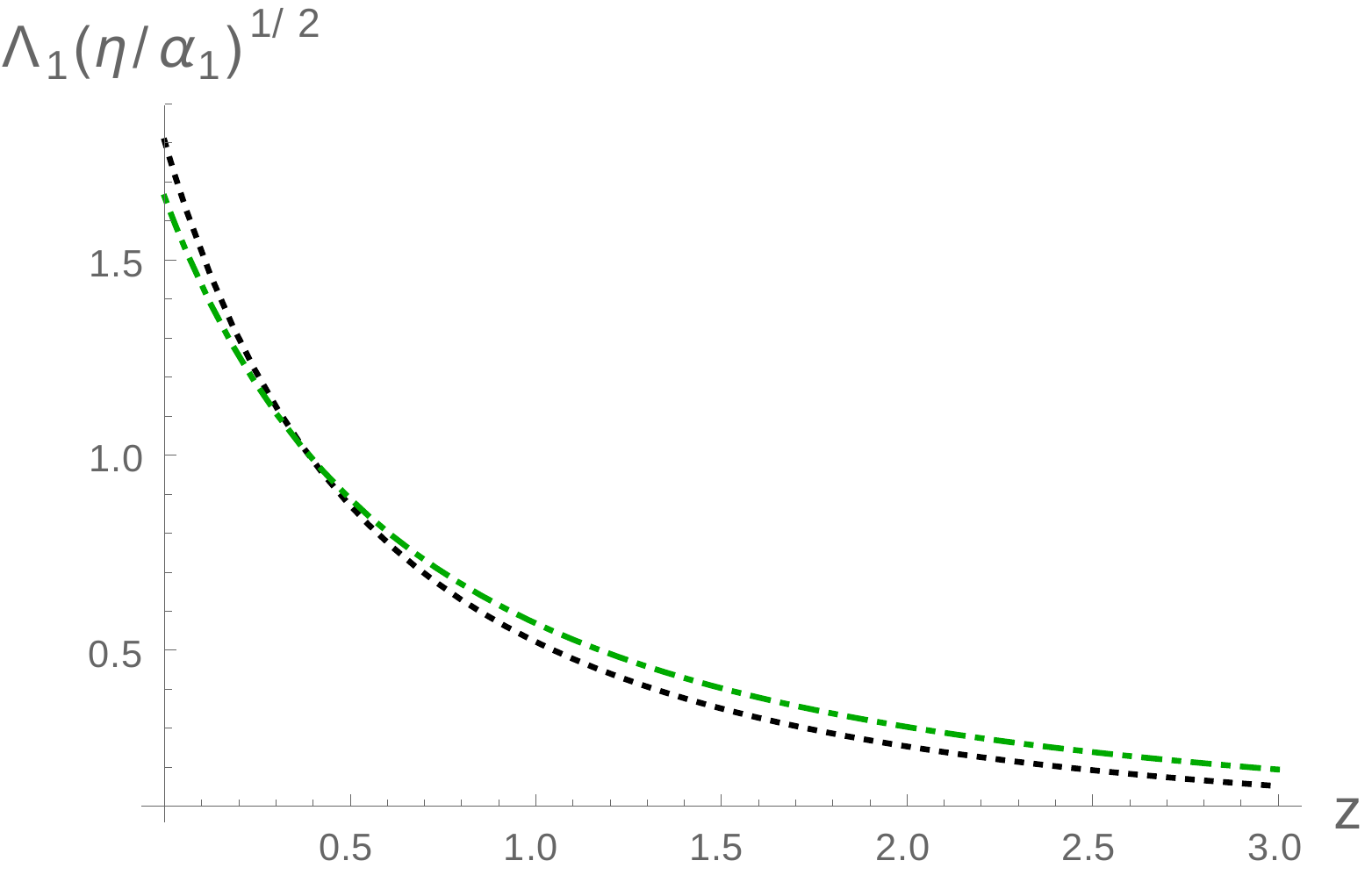}
	\caption{The evolution of the modulated temporal component of the dark energy vector $\Lambda_0$ as a function of the redshift $z$for two non-vanishing values for the constant $\eta=-0.02$ (dot-dahsed) and $\eta=0.1$ (dotted). We use the best fit values for the model parameters presented in \eqref{tab2}.\label{figvec}}
\end{figure}

We have also plotted the temporal component of the dark energy vector field in figure \eqref{figvec} for non-vanishing values of the parameter $\eta$. In the case of vanishing $\eta$, this quantity vanishes as we have discussed earlier. It should be noted that from the structure of the equations, the coupling constant $\alpha$ does not appear alone in the field equations. The only appearace of this constant is in the expression for the temporal component of the dark energy vector field \eqref{vec2}. So, in order to plot the temporal component of the dark energy vector, we have modulated $\Lambda_1$ by $(\alpha_1/\eta)^{1/2}$. It can be seen from the figure that the dark energy vector field tends to zero as the redshift increases. The maximum value of the vector occurs at present time with $z=0$. Also, it should be mentioned that the qualitative behavior of the vector field is the same for different but non-vanishing values of $\eta$.
\begin{table*}
	\begin{center}
		\begin{tabular}{|c||c||c||c||c|} 
			\hline
			~~~~~~~&~~~~~~~~~~~~~~~~~~~~~~$(\Omega_m,\Omega_r)$~~~~~~~~~~~~~~~~~~~~~&~~~~~~~~~$\omega_{eff}$~~~~~~~~~&~~~~~type~~~~~&~~~~~~behavior~~~~~~
			\\
			\hline
			\hline
			$M$&$\left(\f{2(1-\beta_1)}{2+\beta_2},0\right)$&$\f{\beta_2}{2+\beta_2}$&Saddle&dust-like\\
			\hline
			$R$&$\left(0,1-\beta_1\right)$&$\f13(1+4\beta_2)$&Unstable&radiation-like\\
			\hline
			$\Lambda$&$\left(\Omega_m,\f{\Omega_m(2+2\beta_2-\eta\beta_2)+2\eta(\beta_1-1)}{2(\eta-1)\beta_2-2}\right)$&$-1$&Stable&de Sitter\\
			\hline
		\end{tabular}
	\end{center}
	\caption{The fixed points of the dynamical system \eqref{dyna1} and \eqref{dyna2} together with eigenvalues and also their behaviors.\label{tab1}}
\end{table*}
\section{Dynamical system analysis}\label{dyna}
Let us rewrite the friedmann equation \eqref{frid5} in the form
\begin{figure*}
	\includegraphics[scale=0.38]{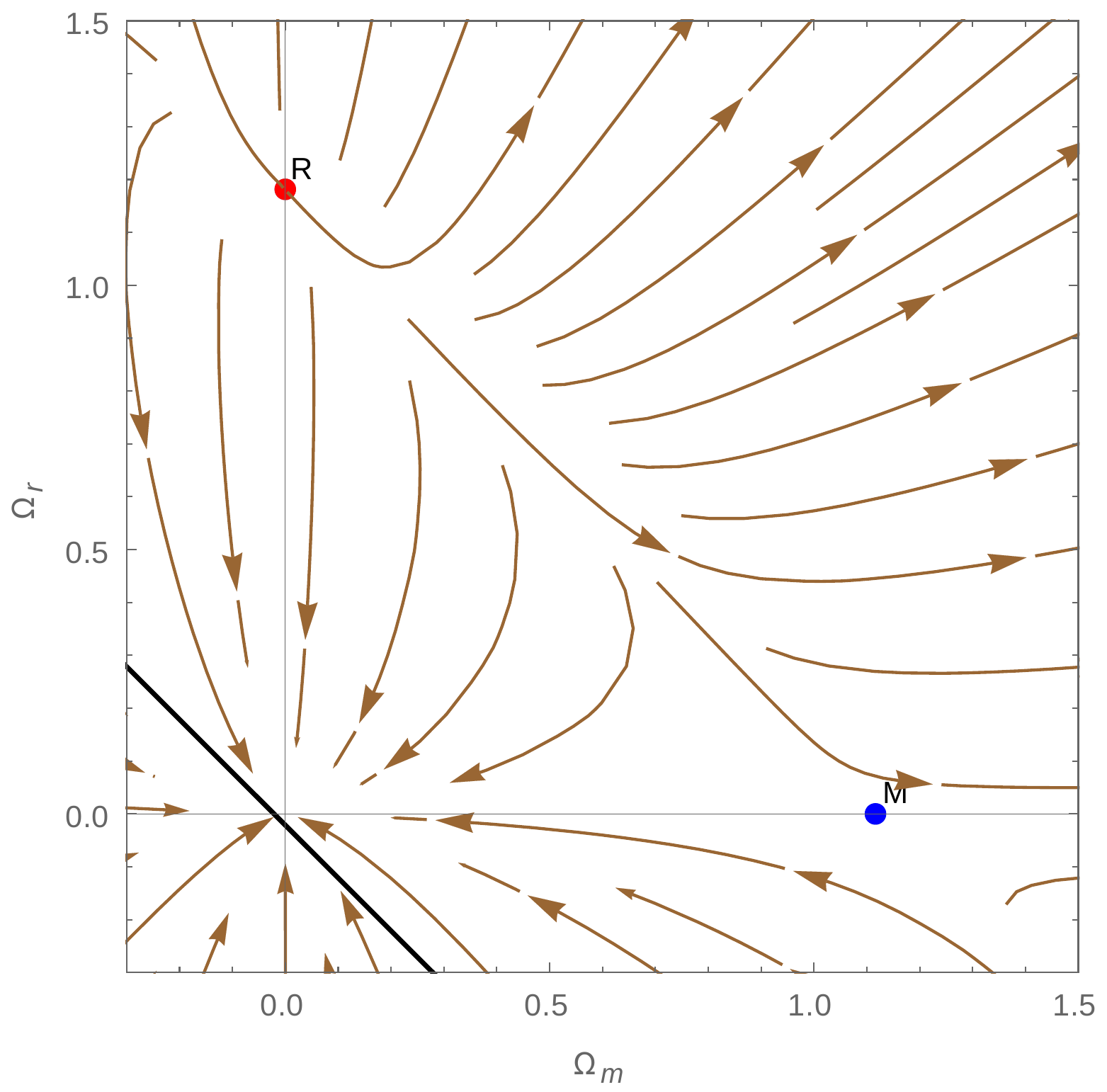}\includegraphics[scale=0.38]{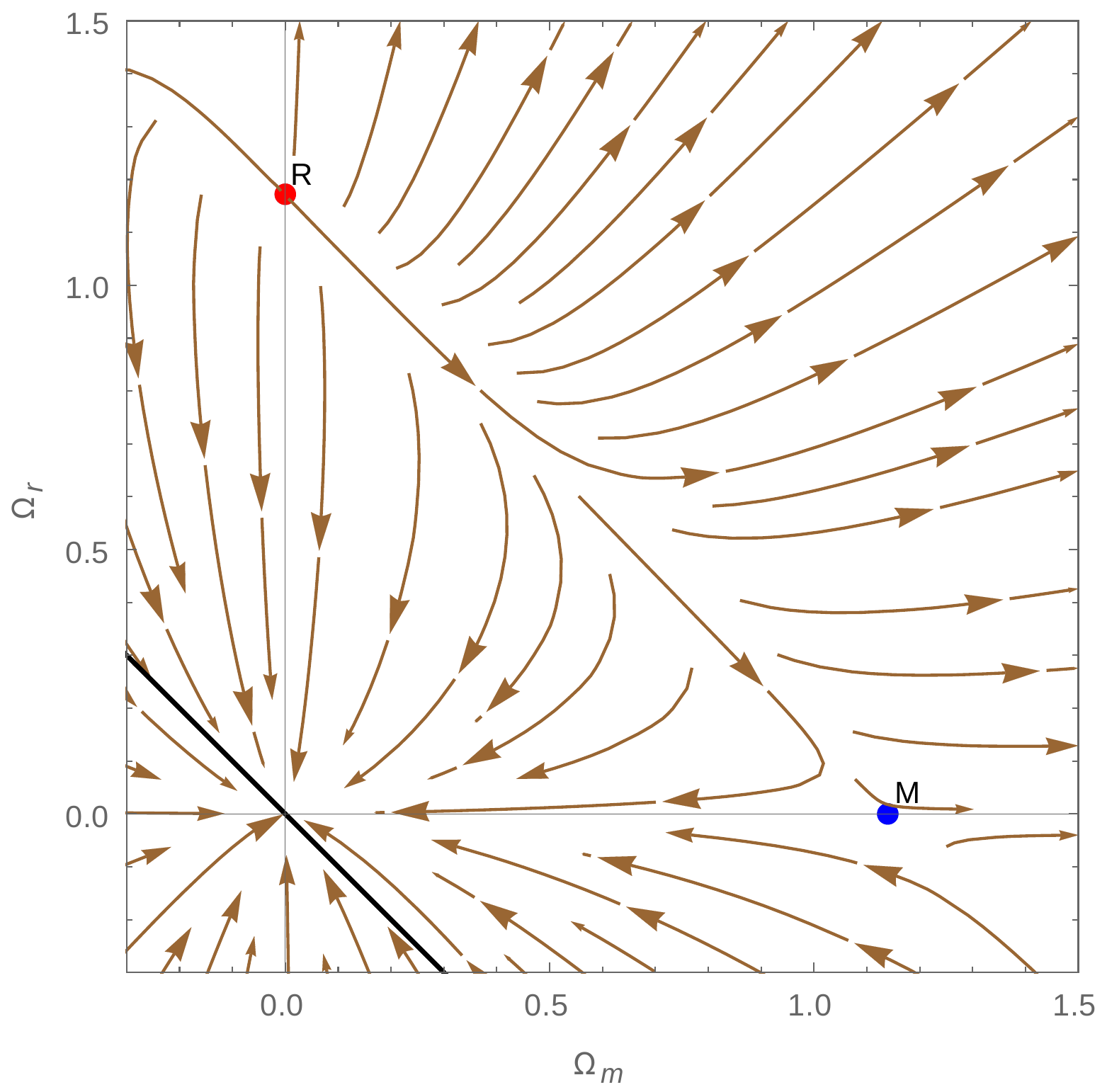}\includegraphics[scale=0.38]{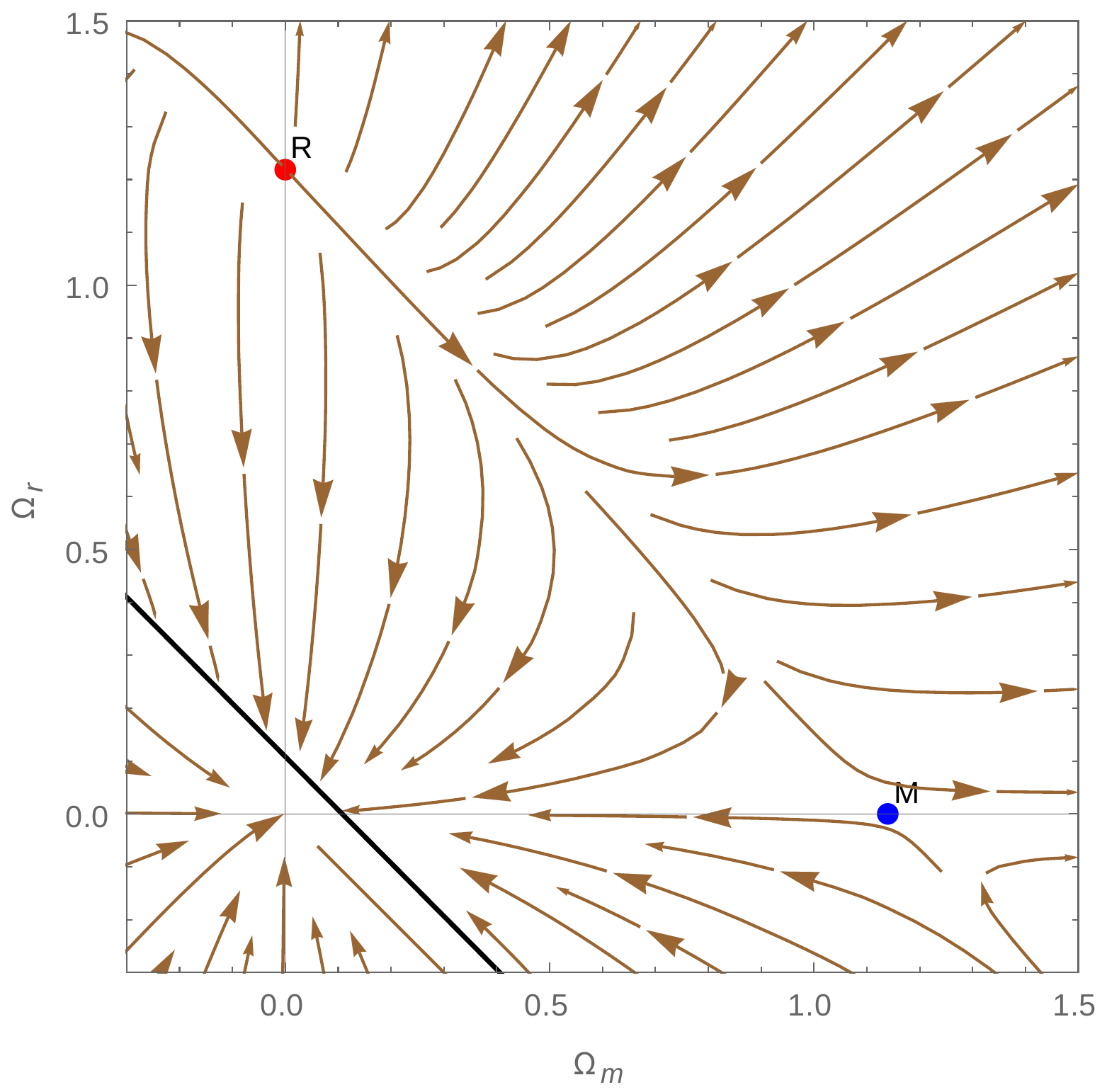}
	\caption{The phase space portrait of the dynamical system \eqref{dyna1} and \eqref{dyna2}. We have plotted the $(\Omega_m,\Omega_r)$ planes for three different values of $\eta=-0.02$ (left), $\eta=0$ (center) and $\eta=0.1$ (right). The corresponding fixed points are also shown. \label{figsys}}
\end{figure*}
\begin{align}
&\left[\Omega_\lambda\left(\f{\eta}{\alpha_1}\right)^{\eta}\f{a^2}{h^2}\right]^{\f{1}{1-\eta}}=\f{1}{\eta-1}\left[\f{a^2}{h^2}(\bar{\rho}_m+\bar\rho_r)\right]^{\f{\eta}{\eta-1}}\nonumber\\&\quad\quad\left[\beta_1-1+\left(1+\f12\beta_2\right)\left(\f{\bar\rho_m a^2}{h^2}\right)+2\left(\f{\bar\rho_r a^2}{h^2}\right)\right].
\end{align}
The above equation suggests that the theory could be analyzed by two dynamical variables
\begin{align}
\Omega_m=\f{\bar\rho_m a^2}{h^2},\qquad \Omega_r=\f{\bar\rho_r a^2}{h^2},
\end{align}
which are the standard dust and radiation densty abundances. Using the conservation equations \eqref{con1} and \eqref{con2}, one can write the dynamical system of the model as
\begin{widetext}
\begin{align}\label{dyna1}
\Omega_m^\prime=\f{\Omega_m}{2}\Bigg(-2+f(\Omega_m,\Omega_r)+\f{6\beta_2(\eta-1)(\Omega_m+\Omega_r)}{(2+\beta_2)(\Omega_m+\Omega_r)-\eta(2-2\beta_1+\beta_2\Omega_r)}\Bigg),
\end{align}
\begin{align}\label{dyna2}
\Omega_r^\prime=\f{\Omega_r}{2}\Bigg(-4+f(\Omega_m,\Omega_r)+\f{6\beta_2(\eta-1)(\Omega_m+\Omega_r)}{2)(\Omega_m+\Omega_r)-\eta(2-2\beta_1-\beta_2\Omega_m)}\Bigg),
\end{align}
where prime represents derivative with respect to $\ln a$ and we have defined
\begin{align}
f(\Omega_m,\Omega_r)=\f{2(\Omega_m+\Omega_r)\Big(-2+2\beta_1+3(1+\beta_2)\Omega_m+4(1+\beta_2)\Omega_r\Big)-(\Omega_m+2\Omega_r)\Big(2-2\beta_1+2\beta_2\Omega_m+4\beta_2\Omega_r\Big)}{(\beta_1-1)(\eta-1)(\Omega_m+\Omega_r)}.
\end{align}
The model is a two dimensional autonomous dynamical system. In order to determine the behavior of the universe at the fixed points, we define an effective equation of state parameter
\begin{align}
\omega_{eff}\equiv-\f13-\f23\f{h^\prime}{h^2}=\f{3(\beta_1-1)+(3\Omega_m+4\Omega_r)(1+\beta_2)}{3(\eta-1)(\beta_1-1)}+\f{2\eta(\beta_1-1)\Omega_r-\eta\beta_2(\Omega_m+2\Omega_r)(3\Omega_m+4\Omega_r)}{6(\eta-1)(\beta_1-1)(\Omega_m+\Omega_r)}.
\end{align}
\end{widetext}
The dynamical system \eqref{dyna1} and \eqref{dyna2} has three fixed points which we have summarized in table \eqref{tab1}. The fixed point $M$ behaved like a matter dominated universe if $\beta_2=0$. So, this point behaves a little different from the standard matter dominated epoch in general relativity. This also can be seen from the conservation equation \eqref{con1}. This fixed point is a saddle point. In figure \eqref{figsys}, we have plotted the stream plot of the dynamical system \eqref{dyna1} and     \eqref{dyna2} for three different values of $\eta=-0.02,0,0.1$. We have also shown the fixed points in the figures. 

The fixed point $R$ behaves like a radiation dominated fixed point if $\beta_2=0$. This fixed point is an unstable fixed point which plays a role of an approximate radiation dominated phase in our model.
Remembering that the dynamical variables $\Omega_m$ and $\Omega_r$ are positive, one can see that these two fixed points behaves as unstable nodes in the positive triangle of the phase portrait. 

The last fixed point $\Lambda$ has an effective equation of state parameter $\omega_{eff}=-1$ and is a de Sitter fixed point. This node is in fact a fixed line as can be seen in figures \eqref{figsys} as a solid black curve. In the case of $\eta=0$, only the point $(\Omega_m,\Omega_r)=(0,0)$ lies in the positive quarter. This point is in fact the stable de Sitter fixed point of the standard $\Lambda$CDM theory. In the case of $\eta=0.1$ we have a fixed line in the positive quarter and one can see from the figure that all the curves end up at this line. We should note that all the points in the line is in fact a fixed point. As a result in this case the phase space is smaller than that of general relativity since the curves can not escape the fixed line to end up at $(0,0)$ point. The case of $\eta=-0.02$ is a little different since no points of the fixed line lie in the positive quarter of the phase space. In this case all the streams in this quarter will end up at the origin which is not a fixed point but as one can see from the figure that is very close to it.

In summary, in all three cases, the evolution of the universe can be started from the radiation dominated fixed point which continue to the matter dominated fixed point and then ends up at a stable de Sitter epoch. As a result the thermal history of the universe can be explained in this model.
\section{Matter perturbation of the model}\label{pert}
 \begin{table*}
	\begin{center}
		\begin{tabular}{|c||c|c|c|c|} 
			\hline
			&Parameter&\qquad Best fit value~~~~~  &\qquad\qquad$1\sigma$ confidence level\qquad\qquad~&\qquad\qquad$2\sigma$ confidence level\qquad\qquad~\\ \hline
			\multirow{4}{*}{$\eta=0.1$}&$\beta_1$&$-0.22$&$-0.28<\beta_1<-0.15$&$-0.34<\beta_1<-0.09$\\ \hhline{|~||*{4}{|-|}}
			&$\beta_2$&$0.14$&$0.11<\beta_2<0.17$&$-0.07<\beta_2<0.20$\\ \hhline{|~||*{4}{|-|}}
			&$\sigma_8^0$&$0.66$&$0.64<\sigma_8^0<0.67$&$0.62<\sigma_8^0<0.69$\\ \hhline{|~||*{4}{|-|}}
			&$\xi$&$1.33$&$0.20<\xi<2.46$&$-0.88<\xi<3.55$\\ \hline\hline
			\multirow{4}{*}{$\eta=0$}&$\beta_1$&$-0.17$&$-0.22<\beta_1<-0.12$&$-0.27<\beta_1<-0.07$\\ \hhline{|~||*{4}{|-|}}
			&$\beta_2$&$0.05$&$0.01<\beta_2<0.08$&$-0.02<\beta_2<0.12$\\ \hhline{|~||*{4}{|-|}}
			&$\sigma_8^0$&$0.76$&$0.74<\sigma_8^0<0.78$&$0.72<\sigma_8^0<0.80$\\ \hhline{|~||*{4}{|-|}}
			&$\xi$&$1.33$&$-9.2<\xi<11.88$&$-19.35<\xi<22.02$\\ \hline\hline
			\multirow{4}{*}{$\eta=-0.02$}&$\beta_1$&$-0.18$&$-0.26<\beta_1<-0.10$&$-0.34<\beta_1<-0.02$\\ \hhline{|~||*{4}{|-|}}
			&$\beta_2$&$0.11$&$0.08<\beta_2<0.15$&$-0.04<\beta_2<0.18$\\ \hhline{|~||*{4}{|-|}}
			&$\sigma_8^0$&$0.66$&$0.64<\sigma_8^0<0.68$&$0.63<\sigma_8^0<0.69$\\ \hhline{|~||*{4}{|-|}}
			&$\xi$&$1.33$&$0.21<\xi<2.4$&$-0.86<\xi<3.53$\\ \hline
		\end{tabular}
	\end{center}
	\caption{The best values of the model parameters $\beta_1$ and $\beta_2$, together with the best values of $\sigma_8^0$ and also the initial condition $\xi$, for three different values of the parameter $\eta=-0.02,0,0.1$.\label{tab2}}
\end{table*}
In this section we will consider the growth of matter perturbation of the Einstein dark energy model. We will consider the scalar perturbations of the field equations in the Newtonian gauge. In this gauge, the scalar perturbation variables $E$, $B$ vanishes and the  perturbed conformal FRW universe can be written as
\begin{align}
ds^2=a^2(t)\Big[-(1+2\varphi)dt^2+(1-2\psi)d\vec{x}^2\Big],
\end{align}
where $\varphi$ and $\psi$ are the Bardeen potentials.
The perturbed energy momentum tensor is defined as
\begin{align}
&\delta T^0_0=-\delta\rho\equiv-\rho\delta,\nonumber\\&\delta T^0_i=(1+w)\rho\partial_i v,\quad\nonumber\\& \delta T^i_j=\delta^i_jc_s^2\rho\delta,
\end{align}
where, $\delta$ is the matter density contrast defined as $\delta=\delta\rho/\rho$, $\rho$ is the background value of the energy density and $v$ is the scalar mode of the velocity perturbation. Also, we have defined the sound speed as $\delta p/\delta\rho=c_s^2$.
The equation of state parameter of the background matter field can be given as $p/\rho=w$. In the following, we will assume that the perturbed and unperturbed matter content of the universe  have the equations of motion of the form $c_s^2=0=w$.

 The scalar mode of perturbed dark energy vector field is taken as
 \begin{align}
 	\Lambda_\mu= a[\Lambda_0+\mathcal{M},\,-\partial_i \mathcal{N} ]
 \end{align}
 where $\Lambda_0=\Lambda_0(t)$ is the non vanishing background component of the vector field $\Lambda_\mu$. 
 
 The  conservation equation \eqref{cons1} up to first order in perturbations can be written as
 \begin{align}
 2(2&+\beta_2-2\alpha\Lambda_0^2)(1+\beta_2-2\alpha\Lambda_0^2)(\theta-3\dot\psi)\nonumber\\&
 -12\beta_2\alpha\Lambda_0 H(\Lambda_0\varphi-\mc{M})+(2+\beta_2-2\alpha\Lambda_0^2)^2\dot\delta=0,
 \end{align}
 and
 \begin{align}
 (2&+\beta_2-2\alpha\Lambda_0^2)\Big[k^2(4\alpha\Lambda_0\mc{M}-\beta_2\delta)+\alpha\Lambda_0\dot\Lambda_0\theta\Big]\nonumber\\&
 +2(2+\beta_2-2\alpha\Lambda_0^2)(1+\beta_2\alpha\Lambda_0^2)(\dot\theta-k^2\varphi)\nonumber\\&
 -4H\Big[\beta_2^2-(1-\alpha\Lambda_0^2)^2\Big]\theta=0,
 \end{align}
 where we have defined $\theta=\nabla_i\nabla^i v$ and Fourier transformed with wave vector $\vec{k}$. Note that in order to simplify the above equations, we have used the background  conservation equation.
 
 From now on, we will work in the sub-horizon limit where the wave number is much greater than the Hubble parameter $k\gg H$. The above equations can be combined to eliminate the variable $\theta$ and we obtain the evolution equation of the matter density contrast, which can be simplified in the sub-horizon limit as
 \begin{align}\label{delta1}
 (2&+\beta_2-2\alpha\Lambda_0^2)\ddot\delta+2\Big[(1-\beta_2-\alpha\Lambda_0^2)H-2\alpha\Lambda_0\dot\Lambda_0\Big]\dot\delta\nonumber\\&+2k^2\Big[(1+\beta_2+\alpha\Lambda_0^2)\varphi+\beta_2\delta-2\Lambda_0\mc{M}\Big]=0.
 \end{align}
 In order to obtain the relations of the variables $\mc{M}$ and $\varphi$, we will use the Einstein and also the dark energy vector field in the sub-horizon limit.
 
  From non-diagonal  components of the Einstein equation, one obtains
 $\psi=\varphi$. Also the $(0)$ component of the vector field equation in sub-horizon limit reads
 \begin{align}
 \mc{M}=\dot{\mc{N}}-H\mc{N}.
 \end{align}
 The $(i)$ component of the vector field equation can be simplified in the sub-horizon limit to
 \begin{align}\label{vecc2}
 H\dot{\mc{N}}+\left[\beta_3\eta\left(\f{\Lambda_0}{\kappa}\right)^{2\eta}\Lambda_0^{-2}a^2+H^2-\alpha a^2\rho\right]\mc{N}=0,
 \end{align}
 which gives the dynamical equation for the variable $\mc{N}$.
The $(00)$ component of the Einstein field equation in the sub-horizon limit can be written as
\begin{align}\label{ein1}
8(1-\beta_1)\kappa^2 k^2\varphi=(2+\beta_2+2\alpha\Lambda_0^2)\rho a^2\delta.
\end{align}
Now, substituting the variable $\varphi$ from the Einstein equation \eqref{ein1} into equation  \eqref{delta1}, one obtains
\begin{align}\label{delta2}
(2&+\beta_2-2\alpha\Lambda_0^2)\ddot\delta+2\Big[(1-\beta_2-\alpha\Lambda_0^2)H-2\alpha\Lambda_0\dot\Lambda_0\Big]\dot\delta\nonumber\\&
+\left[2\beta_2 k^2+\f{(1+\beta_2+\alpha\Lambda_0^2)(2+\beta_2+2\alpha\Lambda_0^2)}{2(\beta_1-1)\kappa^2}a^2\rho\right]\delta\nonumber\\&
+4\kappa^2\alpha\Lambda_0(\dot{\mc{N}}-H\mc{N})=0.
\end{align}
Making equation \eqref{vecc2} dimensionless and substituting the background variables from equation \eqref{vec2}, one obtains
\begin{align}
4h(z)^2\Big[(1+z)\mc{N}(z)^\prime-\mc{N}(z)\Big]=0,
\end{align}
where we have transformed to the redshift coordinates and prime represents derivative with respect to $z$. The above equation has a solution
\begin{align}
\mc{N}(z)=c_1(1+z),
\end{align}
where $c_1$ is an integration constant. One can see that the $\mc{N}$ dependency in equation \eqref{delta2} disappears and one obtains the evolution equation of the density contrast in dimensionless form as
\begin{widetext}
\begin{align}
(2+\beta_2-2\alpha_1\Lambda_1^2)\delta^{\prime\prime}+2\Big[(1-\beta_2-\alpha_1\Lambda_1^2)h-2\alpha_1\Lambda_1\Lambda_1^\prime\Big]\delta^\prime
+\left[2\beta_2 \gamma^2+\f{3(1+\beta_2+\alpha_1\Lambda_1^2)(2+\beta_2+2\alpha_1\Lambda_1^2)}{2(\beta_1-1)}a^2\bar\rho_m\right]\delta=0.
\end{align}
\end{widetext}
It should be noted that in the case of vanishing $\beta_1$, $\beta_2$ and $\alpha$, the above equation reduces to the standard equation of the matter density contrast. In the above equation, we have also defined $\gamma=k/H_0$, which is the dimensionless counterpart of the wave number.

Let us solve the equation governing the evolution of the density contrast. This should be solved together with background equations \eqref{vec2}, \eqref{con1} and \eqref{frid5}. We will use a generalized $\Lambda$CDM initial conditions in deep matter dominated era in which
\begin{align}
\f{d\delta}{d\ln a}|_{z_\star}=\xi\delta|_{z_\star},
\end{align}
where $z_\star$ is some point in the deep matter dominated era which we will assume to be $z_\star=7.1$. Also $\xi$ is a constant which determines deviation from $\Lambda$CDM model. The case $\xi=1$ corresponds to the $\Lambda$CDM model.
\begin{figure}[h]
	\includegraphics[scale=0.45]{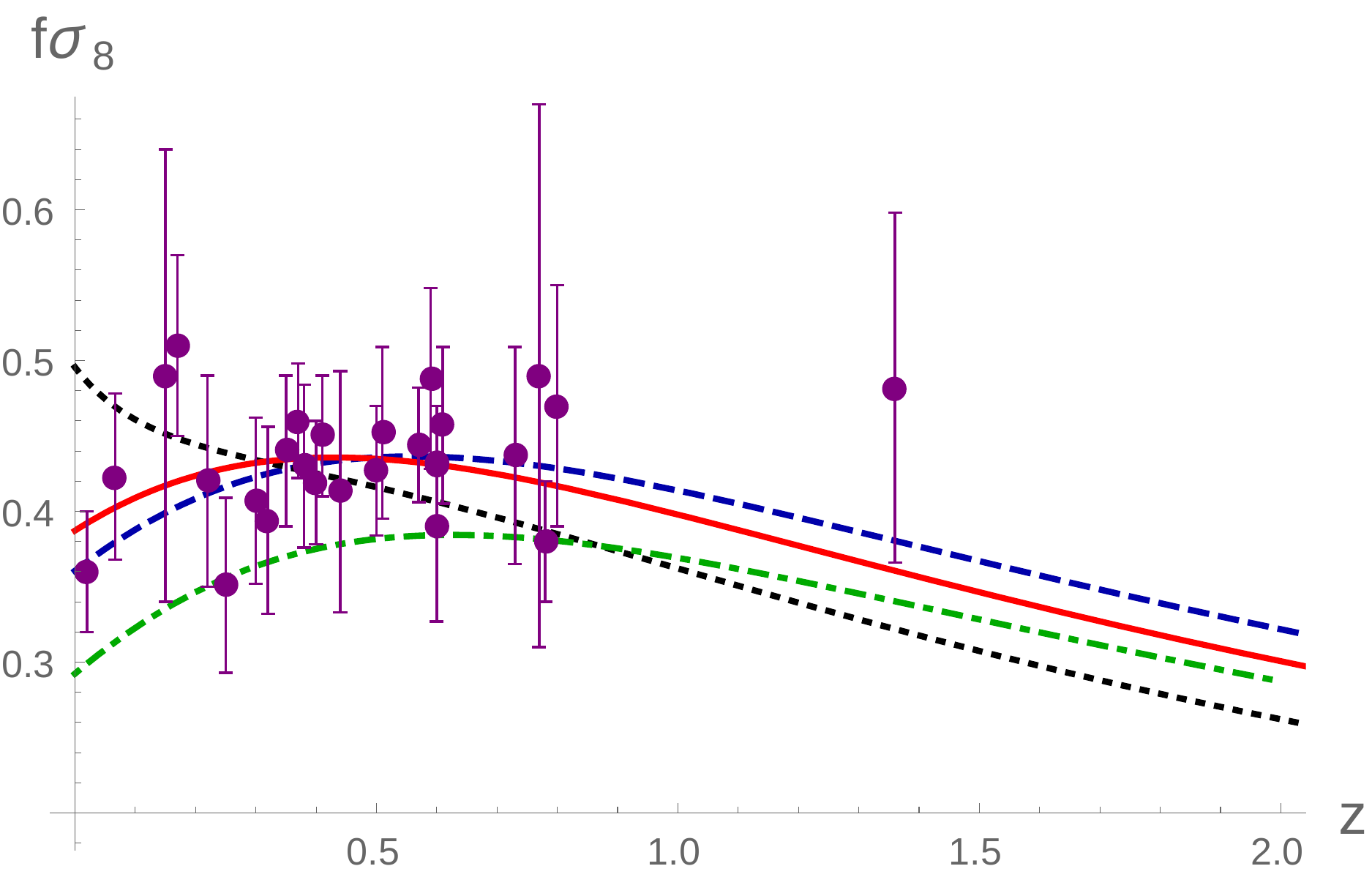}
	\caption{The evolution of $f\sigma_8(z)$ as a function of the redshift $z$ for three different values for the constant $\eta=-0.02$ (dot-dahsed), $\eta=0$ (dashed) and $\eta=0.1$ (dotted). We use the best fit values for the model parameters presented in \eqref{tab2}. The $\Lambda$CDM curve is depicted as a red curve.\label{figfsimga8}}
\end{figure}
To compare the Einstein dark energy model with observational data, we will use two independent data sets on the Hubble parameter in the redshift range $z\sim(0,2)$ \cite{hubble} and also the observational data on $f\sigma_8$ \cite{fsigma8} which is defined as
\begin{align}
f \sigma_8 \equiv \sigma_8(z)f(z).
\end{align}
Here $\sigma_8(z)=\sigma_8^0\, \delta(z)$. The constant $\sigma_8^0$ is model dependent. The growth rate of matter perturbations is defined as
\begin{align}
f=\f{d \ln\delta}{d\ln a}=-(1+z)\f{\delta'}{\delta}.
\end{align}
We estimate the values of $\sigma_8^0$, $\xi$ and also the model parameters $\beta_1$ and $\beta_2$ by maximizing the likelihood function defined as
\begin{align}
\mathcal{L}=\mathcal{L}_{0}e^{-\chi^2/2},
\end{align}
where $\mathcal{L}_0$ is the normalization constant and the quantity $\chi^2$ in our case  is given by
\begin{align}
\chi^2&=\chi_H^2+\chi_{f\sigma_8}^2\nonumber\\&=\sum_i\left(\f{H_{i,obser}-H_{i,theory}}{\sigma_i}\right)^2\nonumber\\&~~~~+\sum_j\left(\f{f\sigma_{8j,obser}-f\sigma_{8j,theory}}{\sigma_j}\right)^2,
\end{align}
where  $H_{i,theory}$ and  $f\sigma_{8j,theory}$ are the theoretical values  for the observables $H_{i,obser}$ and $f\sigma_{8j,obser}$ and $\sigma_i$ is the error of the $i$th data. The total likelihood function is the multiplication of individual likelihoods of the two sets since the data sets we are using here are independent.

In table \eqref{tab2}, we have summarized the best fit values of the parameters $\beta_1$, $\beta_2$, $\xi$ and $\sigma_8^0$ together with their $1\sigma$ and $2\sigma$ confidence intervals for three different values of the parameter $\eta=-0.02,0,0.1$. We have used the best fit values of table \eqref{tab2} to plot the figures. It can be seen from the table that the value of the parameter $\beta_1$ at best fit and also up to $2\sigma$ confidence level is negative. Also, the value of the parameter $\beta_2$ is positive at best fit value. In figure \eqref{figfsimga8}, we have plotted the evolution of the function $f\sigma_8$ as a function of redshift for three different values of the parameter $\eta=-0.02,0,0.1$. We have also plotted the $\Lambda$CDM curve as a red solid line. One can see from the figure that the evolution of the function $f\sigma_8$ for $\eta=0$, is very similar to the $\Lambda$CDM theory. The difference can be traced back to the presence of the term proportional to $\beta_2$. However, for non-vanishing cases of $\eta$ the behavior of this function differs significantly from the $\Lambda$CDM theory. However, all cases satisfied observational data. The present value of $f\sigma_8$ for positive values of the parameter $\eta$ is greater than that of the $\Lambda$CDM value and for negative values of $\eta$ its value becomes smaller. One can then see that for redshifts greater than $0.2$, non-vanishing values of $\eta$ fits very well with observations. In summary, it sees that more data would be needed to decide the best model which fits the data.

\section{Conclusion and final remarks}
In this paper we have considered cosmological implications of the Einstein dark energy model. This model consists of an Einstein-Hilbert Lagrangian with a coupling proportional to the trace of the energy-momentum tensor, coupled to a dark energy vector field. The vector field has a non-minimal coupling with matter fields, which make the energy-momentum tensor non-conservative. This non-conservation of the energy-momentum tensor results in creation of matter out of geometry. The rate of such a creation is calculated in \cite{EDE}. We have also considered a power-law potential term for dark energy vector field.

We have obtained the best fit values of the model parameters by using two sets of independent data from Hubble parameter and also the function $f\sigma_8$. At redshifts smaller than $z\sim3$, the behavior of the Hubble parameter is very similar to the $\Lambda$CDM model for small values of the paramter $\eta$. For large values of this parameter, the behavior of the Hubble parameter is very different and one can not find a best fit with observational data. As a result larger values of $\eta$ is ruled out by observations and so we have considered small values in this paper. Despite the behavior of the Hubble parameter the matter density abundance behaves differently from $\Lambda$CDM model. At larger redshifts, the matter density abundance is larger than the $\Lambda$CDM value implying that there are more matter present at those redshifts. However, the matter density decreases more rapid than $\Lambda$CDM model implying that the present values of the matter density abundance is the same as $\Lambda$CDM value. This shows that the rate of changing matter content of the universe to curvature is getting smaller at late times. This can also be seen from the evolution of the function $f_0$ in figure \eqref{figvec}, since the rate of creation of matter is proportional to this function \cite{EDE}. 

Dynamical system analysis of the theory can also show this behavior. As we have discussed in this paper, the theory has three fixed points, one of them is an exact de Sitter node. However, We have two fixed points which behave a little different from $\Lambda$CDM  matter and radiation nodes, since the constant $\beta_2$ is small. The difference is directly related to the term $\beta_2 T$ in the action which comes from Rastall's idea. This makes the theory non-conservative and consequently the fixed points become different. However, the thermal history of the universe can be achieved in this theory since all the required fixed points are present.

We have also considered the first order perturbation analysis of the model and obtain the growth rate of matter density contrast in the sub-horizon limit. The differential equation governing the behavior of matter densit contrast is affected by both Rastall's term $\beta_2T$ and also by non-minimal coupling term $\mc{L}_m\Lambda_\mu\Lambda^\mu$ term. Since the theory is different from $\Lambda$CDM theory at every times, we have modified the initial conditions on the density contrast to cover this new model. From the best fit values presented in table \eqref{tab2}, one can see that at deep matter dominated epoch, the rate of change of $\delta$ is bigger than that of $\Lambda$CDM theory by about $30\%$. However, the value of $\sigma_8^0$ in this theory does not change much compared to $\Lambda$CDM model. We have also shown that the Einstein dark energy model can be compatible with observational data for redshifts greater than $0.2$. 

In summary we would like to say that the Einstein's idea of considering elementary particles as electromagnetic like fields can be put forward to make some progress in obtaining satisfactory model of dark energy. But more data would still be needed to decide which model is more friendly with observations.

\end{document}